\documentclass[%
 reprint,
 amsmath,amssymb,
 aps,
]{revtex4-1}

\usepackage{graphicx}
\usepackage{dcolumn}
\usepackage{bm}

\begin{document}

\preprint{APS/123-QED}

\title{How a spinning quark moves in the NJL type mean field}

\author{Feng Li}
\email{fengli@fias.uni-frankfurt.de}
\date{\today}

\begin{abstract}
The transport equation of the spinning quarks, moving in the Nambu-Jona-Lasinio (NJL) type mean field, is derived by solving the Schwinger-Dyson equations, up to the order of $\hbar$, under the mean field approximation. It shows that the scalar and the vector mean field potentials have different impacts on the quantum correction to the transport equation. Both the forces give rise to an anomalous velocity and an anomalous force where the latter would vanish if the vector force is turned off. The particle dispersion relation is modified by the vector force as well. Besides, the spin precession in the particle rest frame is purely governed by the vector force.
\end{abstract}

\pacs{Valid PACS appear here}
\maketitle


\section{\label{sec:intro}Introduction}
Great attention in the community of the high energy nuclear physics is drawn to the motion of the spinning particles. The equations of motion of the chiral particles moving in a permanent magnetic field is derived\cite{Stephanov:2012ki,Son:2012wh,Son:2012zy,Gao:2012ix,Chen:2012ca,Manuel:2014dza,Chen:2015gta,Hidaka:2016yjf,Huang:2018wdl}. And within these equations, an anomalous term corresponding to the chiral magnetic effect\cite{Kharzeev:2004ey,Kharzeev:2007tn,Kharzeev:2007jp,Fukushima:2008xe} is discovered and regarded as a probe of instanton and the QCD vacua\cite{GrossPisarskiYaffe1981}. Meanwhile, quarks are also affected by the strong forces which can be modelled by the Nambu-Jona-Lasinio (NJL) Lagrangian \cite{Nambu:1961fr,Nambu:1961tp}. Under the mean field approximation, the NJL Lagrangian includes both a scalar and a vector mean field potential\cite{Hatsuda:1994pi,Buballa:2003qv} where the latter one is coupled with the quark field in the same way as the electromagnetic field does. For a baryon rich rotating fireball, the scalar mean field potential provides an attractive force pointing to the center of the fireball, while the vector one provides both a repulsive electric-like force pointing away from the center, and a magnetic-like field pointing along the rotating axis. Given the strength of the strong force, its anomalous effect on the motion of the spinning quarks should be at least as important as the one contributed by the magnetic field and should be derived and taken into consideration when we carry out simulations to study the anomalous effect.  

On the other hand, the existence of the vector interactions among the quarks is still controversial. It is excluded by the comparison with the equations of the state obtained in the Lattice QCD calculation\cite{Steinheimer:2010sp}, but is necessary for explaining the existence of the neutron star of two solar mass\cite{Klahn:2013kga,Baym:2017whm}. The vector interaction is also considered as a mechanism leading to the different elliptical flows of the particles and anti-particles\cite{Song:2012cd,Xu:2013sta}. We will see, in the following sections, that the scalar and vector mean field potentials have different impacts on the quantum correction to the equations of motion of the spinning quarks. Both give rise to an anomalous velocity and an anomalous force where the latter vanishes if the vector force is turned off. The dispersion relation is modified by the vector force as well. Besides, the spin precession in the particle rest frame is governed by the vector force only. We therefore expect the motions of the spinning quarks to be very different, in a baryon rich rotating fireball, in the cases with and without the vector mean field potential and thus to provide a new perspective to study the existence of the vector interactions among the quarks.

This work is inspired by a recently published paper which provides a complete and consistent chiral transport from Wigner function formalism\cite{Huang:2018wdl}. Most derivations are similar. But instead of starting from a quantum kinetic equation\cite{Vasak:1987um} derived for the U(1) gauge theory, we start our derivations from the Schwinger-Dyson equations\cite{Rammer:2007zz} which have already been employed for deriving a full classical transport equation for the NJL model\cite{Klevansky:1997wm}. The next sections are organized as follows. In section II, we obtain 32 primary equations, grouped in 10 groups, by truncating the Schwinger-Dyson equations at the order of $\partial_p$ under the mean field approximation. These equations are solved at the order of $\hbar^0$ in section III where the classical transport equation is obtained, and at the order of $\hbar$ in section IV where the equation describing the spin precession and the quantum correction to the transport equation are obtained. The conclusions and outlooks are in section V.

\section{\label{sec:primary}Primary Equations}
The Schwinger-Dyson equations are
\begin{eqnarray*}
G(x_1, x_1^\prime) &=& G_0(x_1, x_1^\prime)\\
&+& \int dx_2 \int dx_3 G_0(x_1, x_2)\Sigma(x_2, x_3) G(x_3, x_1^\prime), \nonumber\\
G(x_1, x_1^\prime) &=& G_0(x_1, x_1^\prime)\\
&+& \int dx_2 \int dx_3 G(x_1, x_2)\Sigma(x_2, x_3) G_0(x_3, x_1^\prime),
\end{eqnarray*}
where $G(x,y)$ is the two point Green's function, $\Sigma(x,y)$ is the self energy, and $G_0(x,y)$ is the bare Green's function which fulfills
\begin{equation}
    \hat G^{-1}_0(x) G_0(x,y) = G_0(x,y)\hat G^{\prime -1}_0(y)= i\delta(x-y)\sigma_3,
\end{equation}
with
\begin{eqnarray}
\hat G^{-1}_0(x) &\doteq& i\hbar{\partial\mkern-11 mu/}_x - m_0, \nonumber\\
\hat G^{\prime -1}_0(y) &\doteq& -i\hbar\overleftarrow{\partial\mkern-11 mu/}_y - m_0.
\end{eqnarray}
The above Green's functions, self energies, and the third Pauli matrix $\sigma_3$, are defined in the so-called Schwinger-Keldysh space \cite{Schwinger:1960qe, Keldysh:1964ud} where the Green's function is written as
\begin{equation*}
    G = \left(
    \begin{array}{cc}
      G_F   &  G_> \\
      G_<   &  G_{AF}
    \end{array}
    \right)
\end{equation*} with
\begin{eqnarray*}
G_>(x,y) &\doteq& \langle \psi (x) \bar\psi(y)\rangle \nonumber\\
G_<(x,y) &\doteq& -\langle \bar\psi (y) \psi(x)\rangle \nonumber\\
G_F(x,y) &\doteq& \theta(y^0-x^0) G_> (x, y) + \theta(x^0 - y^0) G_<(x,y) \nonumber\\
G_{AF}(x,y) &\doteq& \theta(y^0-x^0) G_< (x, y) + \theta(x^0 - y^0) G_>(x,y).
\end{eqnarray*}
Under the mean field approximation, the self energy $\Sigma(x,y)$ is reduced to $iU(x)\delta(x-y)\sigma_3$\cite{Klevansky:1997wm}, where the mean field potential $U(x)$ can be decomposed into a scalar and a vector part, i.e.,
\begin{equation}
    U(x) = \Phi(x) + \gamma^\mu \Omega_\mu(x).
\end{equation}
The Dyson equations are thus reduced to
\begin{eqnarray}
\label{eq:Dyson1}
G(x_1, x_1^\prime) &=& G_0(x_1, x_1^\prime) \nonumber\\
&&+ i\int dx_2 G_0(x_1, x_2)U(x_2)\sigma_3 G(x_2, x_1^\prime),\\
\label{eq:Dyson2}
G(x_1, x_1^\prime) &=& G_0(x_1, x_1^\prime) \nonumber\\
&&+ i\int dx_2G(x_1, x_2)U(x_2)\sigma_3 G_0(x_2, x_1^\prime).
\end{eqnarray}
After applying $\hat G^{-1}(x_1)$ and $\hat G^{\prime -1}(x_1^\prime)$ to Eq.(\ref{eq:Dyson1}) and Eq.(\ref{eq:Dyson2}) respectively, and taking the difference between the diagonal elements in the Schwinger-Keldysh space, we obtain
\begin{eqnarray}
\label{eq:schodinger1}
(i\hbar{\partial\mkern-11 mu/}_{x_1}-m_0+U(x_1))G_K(x_1, x_1^\prime) &=& 0, \\
\label{eq:schodinger2}
G_K(x_1, x_1^\prime)(-i\hbar{\partial\mkern-11 mu/}_{x_1^\prime}-m_0+U(x_1)) &=& 0,
\end{eqnarray}
where $G_K \doteq (G_> + G_<)/2 = (G_F + G_{AF})/2$ is the so-called Keldysh Green's function.
After taking the sum and difference between Eq.(\ref{eq:schodinger1}) and Eq.(\ref{eq:schodinger2}), substituting $x_1$, $x_1^\prime$ with $X \doteq (x_1 + x_1^\prime) / 2$ and $x \doteq x_1 - x_1^\prime$, and applying the Fourier transformation
\begin{eqnarray*}
\widetilde G(X,p) &\doteq& \int d^4x e^{ipx/\hbar}G_K(X+x/2, X-x/2), \\
\widetilde U(p) &\doteq& \int d^4x e^{ipx/\hbar}U(x),
\end{eqnarray*}
\begin{widetext}
we obtain
\begin{eqnarray}
\label{eq:NoApprox1}
\frac i 2\partial_{X^\mu}\{\gamma^\mu,{\widetilde G}(X,p)\}+[{p\mkern-11 mu/},\widetilde G(X,p)]+\int \frac{d^4l}{(2\pi)^4}\left({\widetilde U}(l){\widetilde G}(X,p-\frac l 2)-{\widetilde G}(X,p+\frac l 2){\widetilde U}(l)\right)e^{-ilX/\hbar}=0,\\
\label{eq:NoApprox2}
\frac i 2\partial_{X^\mu}[\gamma^\mu,{\widetilde G}(X,p)]+\{{p\mkern-11 mu/}-m_0,\widetilde G(X,p)\}+\int \frac{d^4l}{(2\pi)^4}\left({\widetilde U}(l){\widetilde G}(X,p-\frac l 2)+{\widetilde G}(X,p+\frac l 2){\widetilde U}(l)\right)e^{-ilX/\hbar}=0.
\end{eqnarray}
Taylor expanding $\widetilde G(X,p+\frac l 2)$ at $l = 0$ as
\begin{equation}
    \widetilde G(X,p+\frac l 2) \approx {\widetilde G}(X,p)\pm \frac{l_\mu}{2}\partial_{p_\mu}{\widetilde G}(X,p)+\mathcal O(l^2),
\end{equation}
we obtain from Eq.(\ref{eq:NoApprox1}) and Eq.(\ref{eq:NoApprox2}) that
\begin{eqnarray}
\label{eq:DysonApprox1}
\frac i 2\hbar\partial_{X^\mu}\{\gamma^\mu,{\widetilde G}(X,p)\}+[{k\mkern-11 mu/},{\widetilde G}(X,p)]-\frac i 2\hbar\{\partial_{X^\mu}U(X),\partial_{p_\mu}{\widetilde G}(X,p)\}+\mathcal O(\hbar^2)=0,\\
\label{eq:DysonApprox2}
\frac i 2\hbar\partial_{X^\mu}[\gamma^\mu,{\widetilde G}(X,p)]+\{{k\mkern-11 mu/}-M,{\widetilde G}(X,p)\}-\frac i 2\hbar[\partial_{X^\mu}{\Omega\mkern-11 mu/}(X),\partial_{p_\mu}{\widetilde G}(X,p)]+\mathcal O(\hbar^2)=0,
\end{eqnarray}
where $k\doteq p+\Omega$ is the kinetic momentum and $M \doteq m_0 - \Phi$ is the effective mass. After we substitute $\widetilde G(X,p)$ with $\widetilde G(X,k)$, the derivatives change as $\partial_{p_\mu} \to \partial_{k_\mu}$, $\partial_{x^\mu} \to \partial_{x^\mu} + \partial_{x^\mu}\Omega_\nu \partial_{k_\nu}$, and Eq.(\ref{eq:DysonApprox1}) and Eq.(\ref{eq:DysonApprox2}) becomes
\begin{eqnarray}
\label{eq:DysonApprox3}
\frac i 2\hbar D_\mu\{\gamma^\mu,{\widetilde G}(X,k)\}+[{k\mkern-11 mu/},{\widetilde G}(X,k)]-\frac i 2\hbar\{\partial_{X^\mu}\Phi(X),\partial_{k_\mu}{\widetilde G}(X,k)\}+\mathcal O(\hbar^2)&=&0,\\
\label{eq:DysonApprox4}
\frac i 2\hbar D_\mu[\gamma^\mu,{\widetilde G}(X,k)]+\{{k\mkern-11 mu/}-M,{\widetilde G}(X,k)\}+\mathcal O(\hbar^2)&=&0,
\end{eqnarray}
\end{widetext}
where $D_\mu \doteq \partial_{X^\mu} + F_{\mu\nu}\partial_{k_\nu}$, which fulfills
\begin{equation}
\label{eq:Dk}
    D_\mu k_\nu = F_{\mu\nu}
\end{equation}
and
\begin{equation}
\label{eq:DD}
    [D_\mu,D_\nu] = -\partial_{X^\tau}F_{\mu\nu}\partial_{k^\tau},
\end{equation}
with $F_{\mu\nu}$ being $\partial_\mu \Omega_\nu - \partial_\nu \Omega_\mu$. Decomposing $\widetilde G$ as
\begin{equation*}
    \widetilde G(X,k) = S + iP\gamma_5 + V_\mu\gamma^\mu + A_\mu\gamma^\mu\gamma_5 + \frac 1 2 J_{\mu\nu}\sigma^{\mu\nu}
\end{equation*}
with $\sigma^{\mu\nu}$ being $\frac i 2 [\gamma^\mu, \gamma^\nu]$, and using
\begin{eqnarray*}
\{\gamma^\mu,\widetilde G\} &=& 2S\gamma^\mu + 2V^\mu + \epsilon^{\mu\nu\rho\sigma}(A_\nu\sigma_{\rho\sigma}-J_{\rho\sigma}\gamma_\nu\gamma_5)\\{}
 [\gamma^\mu,\widetilde G] &=& 2iP\gamma^\mu\gamma_5 - 2i V_\nu \sigma^{\mu\nu} + 2 A^\mu \gamma_5 + 2i J^{\mu\nu} \gamma_\nu
\end{eqnarray*}
which are derived from the identities of the $\gamma$ matrices, we obtain from Eq. (\ref{eq:DysonApprox3}) and Eq. (\ref{eq:DysonApprox4}) that
\begin{eqnarray}
\label{eq:1scalar}
0 &=& \hbar D_\mu V^\mu-\hbar \partial_{X^\mu}\Phi\partial_{k_\mu}S + \mathcal O(\hbar^3), \\
0 &=& 2A^\mu k_\mu + \hbar\partial_{X^\mu}\Phi\partial_{k_\mu} P + \mathcal O(\hbar^2), \\
0 &=& \hbar D_\mu S + 2k^\nu J_{\nu\mu} -\hbar \partial_{X^\nu}\Phi\partial_{k_\nu} V_\mu + \mathcal O(\hbar^2),\\
0 &=& \frac \hbar 2 \epsilon^{\mu\nu\rho\sigma} D_\mu J_{\rho\sigma} - 2k^\nu P +\hbar\partial_{X^\mu}\Phi\partial_{k_\mu} A^\nu \nonumber\\
&& + \mathcal O(\hbar^2), \\
0 &=& \frac \hbar 2 \epsilon^{\mu\nu\rho\sigma} D_\mu A_\nu - (k^\rho V^\sigma - k^\sigma V^\rho)\nonumber\\
&&-\frac \hbar 2 \partial_{X^\mu}\Phi\partial_{k_\mu} J^{\rho\sigma} + \mathcal O(\hbar^2), \\
0 &=& k_\mu V^\mu - MS + \mathcal O(\hbar^2), \\
0 &=& \hbar D_\mu A^\mu - 2MP + \mathcal O(\hbar^2),\\
0 &=& \hbar D_\mu J^{\nu\mu} + 2k^\nu S - 2MV^\nu + \mathcal O(\hbar^2),\\
0 &=& \hbar D_\mu P + \epsilon_{\rho\sigma\nu\mu} J^{\rho\sigma}k^\nu + 2MA_\mu  + \mathcal O(\hbar^2), \\
\label{eq:2spin}
0 &=& \hbar D_\mu V_\nu - \hbar D_\nu V_\mu +2 \epsilon_{\mu\nu\rho\sigma}k^\rho A^\sigma-2MJ_{\mu\nu}\nonumber\\
&&+ \mathcal O(\hbar^2).
\end{eqnarray}
It might not be that transparent why the higher order term in Eq.(\ref{eq:1scalar}), which is the scalar component of Eq.(\ref{eq:DysonApprox3}), is $\mathcal O(\hbar^3)$ rather than $\mathcal O(\hbar^2)$. It is because the lowest order term in $\mathcal O(\hbar^2)$ in Eq.(\ref{eq:DysonApprox3}) is $-\frac{\hbar^2}{8}[\partial_{X^\mu}\partial_{X^\nu}U,\partial_{p_\mu}\partial_{p_\nu}\widetilde G]$, which does not have a scalar component. Eq.(\ref{eq:1scalar}-\ref{eq:2spin}) are the primary equations which will be solved in the next two sections. To solve them, we further expand $\widetilde G$ as $\widetilde G = G_{(0)} + \hbar G_{(1)} + \mathcal O(\hbar^2)$. The components of $G_{(0)}$ will be solved in the next section.

\section{zeroth order: the classical transport equation}
In this section, we only investigate the components of $G_{(0)}$ and neglect all the terms of the order $\hbar$. Eq.(\ref{eq:1scalar}-\ref{eq:2spin}) are reduced to
\begin{eqnarray}
\label{eq:1scalar0}
D_\mu V_{(0)}^\mu - \partial_{X^\mu}\Phi\partial_{k_\mu}S_{(0)} + \mathcal O(\hbar)&=& 0, \\
\label{eq:1pscalar0}
2A_{(0)}^\mu k_\mu + \mathcal O(\hbar)&=& 0, \\
\label{eq:1vector0}
2k_\nu J^{\nu\mu}_{(0)} + \mathcal O(\hbar)&=& 0,\\
\label{eq:1avector0}
2k^\nu P_{(0)} + O(\hbar) &=& 0, \\
\label{eq:1spin0}
k^\rho V^\sigma_{(0)} - k^\sigma V^\rho_{(0)} + \mathcal O(\hbar) &=&0, \\
\label{eq:2scalar0}
k_\mu V^\mu_{(0)} - MS_{(0)} + \mathcal O(\hbar) &=& 0, \\
\label{eq:2pscalar0}
-2MP_{(0)} + \mathcal O(\hbar) &=& 0,\\
\label{eq:2vector0}
 2k^\nu S_{(0)} - 2MV^\nu_{(0)} + \mathcal O(\hbar) &=& 0,\\
 \label{eq:2avector0}
 \epsilon_{\rho\sigma\nu\mu} J^{\rho\sigma}_{(0)}k^\nu + 2MA_\mu^{(0)}  + \mathcal O(\hbar) &=& 0, \\
\label{eq:2spin0}
 2 \epsilon_{\mu\nu\rho\sigma}k^\rho A^\sigma_{(0)}-2MJ_{\mu\nu}^{(0)}+ \mathcal O(\hbar) &=& 0.
\end{eqnarray}
Some of the above equations are redundant. Both Eq.(\ref{eq:1avector0}) and Eq.(\ref{eq:2pscalar0}) indicate $P_{(0)} = 0$, and Eq.(\ref{eq:1pscalar0}) and Eq.(\ref{eq:1vector0}) carry no more information than Eq.(\ref{eq:2avector0}) and Eq.(\ref{eq:2spin0}). We learn $V^\mu_{(0)} = k^\mu \mathcal F_{(0)}(X,k)$ from Eq.(\ref{eq:1spin0}), and
\begin{eqnarray}
S &=& \frac{k^2}{M}\mathcal F_{(0)}(X,k), \\
\label{eq:F0}
\mathcal F_{(0)}(X,k) &=& \eta^{(0)}(X,k)\delta(k^2-M^2),
\end{eqnarray}
 from Eq.(\ref{eq:2scalar0}) and Eq.(\ref{eq:2vector0}), where $\eta^{(0)}$ is a scalar function. Finally, a general solution to Eq.(\ref{eq:2avector0}) and Eq.(\ref{eq:2spin0}) is
\begin{eqnarray}
\label{eq:A00}
A^\mu_{(0)} &=& -Mn^\mu(X,k)\xi^{(0)}(X,k)\delta(k^2-M^2),\\
\label{eq:J00}
J^{\mu\nu}_{(0)} &=& -\epsilon^{\mu\nu\rho\sigma}k_\rho n_\sigma(X,k)\xi^{(0)}(X,k)\delta(k^2-M^2),
\end{eqnarray}
where $n$ fulfills $n\cdot k = 0$ and $\xi^{(0)}$ is another scalar function. $\eta^{(0)}$ and $\xi^{(0)}$ in Eq.(\ref{eq:F0}), Eq.(\ref{eq:A00}) and Eq.(\ref{eq:J00}) can be replaced with $f^{(0)}_+ + f^{(0)}_-$ and $f^{(0)}_+ - f^{(0)}_-$, respectively, and $V^{(0)}$, $A^{(0)}$ and $J^{(0)}$ can thus be written as
\begin{eqnarray}
V^\mu_{(0)} &=& \sum_{s=\pm} k^\mu f_s^{(0)}(X,k)\delta(k^2-M^2),\\
\label{eq:A0}
A^\mu_{(0)} &=& -\sum_{s=\pm}sMn^\mu(X,k)f_s^{(0)}(X,k)\delta(k^2-M^2),\\
\label{eq:J0}
J^{\mu\nu}_{(0)} &=& -\sum_{s=\pm}s\epsilon^{\mu\nu\rho\sigma}k_\rho n_\sigma(X,k)f_s^{(0)}(X,k)\delta(k^2-M^2),\nonumber\\
\end{eqnarray}
where $f_s^{(0)}$ can be interpreted, at the semi-classical level, as the distribution function of the quarks with spin $s$ which, according to Eq.(\ref{eq:1scalar0}), fulfills the classical transport equation 
\begin{eqnarray}
\label{eq:ClassTransport}
    0&=&\delta(k^2-M^2)\sum_{s=\pm}\left[k^\mu D_\mu f^{(0)}_s - M\partial_{X^\mu}\Phi\partial_{k_\mu}f^{(0)}_s\right]\nonumber\\
    &&+ \mathcal O(\hbar^2).
\end{eqnarray}
The derivatives on $\delta(k^2-M^2)$ does not appear in Eq.(\ref{eq:ClassTransport}) since $k^\mu D_\mu \delta(k^2-M^2) = (2 F_{\mu\nu}k^\mu k^\nu + 2 M k^\mu \partial_{X^\mu} \Phi) \delta^\prime(k^2 - M^2)$ where the first term vanishes and the second term is cancelled with $-M\partial_{X^\mu}\Phi\partial_{k_\mu}\delta(k^2-M^2)$.

The current and axial-current density can be expressed, at the semi-classical level, using $f_s^{(0)}$ as
\begin{eqnarray*}
j^\mu &=& \sum_{s=\pm}\int d^4 k k^\mu f_s^{(0)} \delta(k^2-M^2) + \mathcal O(\hbar),\\
j_5^\mu &=& -\sum_{s=\pm}\int d^4 k sM n^\mu f_s^{(0)} \delta(k^2-M^2) + \mathcal O(\hbar)
\end{eqnarray*}

In summary, the zeroth order of the Green's function $G_{(0)} = \sum_s(1+s\gamma_5{n\mkern-11 mu/})({k\mkern-11 mu/}+M)f_s^{(0)}\delta(k^2-M^2)$, which, according to the textbook on quantum field theory\cite{Srednicki:2007qs}, is equal to $\sum_s u_s(\mathbf k)\bar u_s(\mathbf k) f_s^{(0)}\delta(k^2-M^2)$ where $u_s$ is the Dirac spinor and $f_s^{(0)}$ fulfills Eq.(\ref{eq:ClassTransport}). And by comparing $G_{(0)}$ with the expression in Ref.\cite{Srednicki:2007qs}, we know that $n^\mu = \mathcal B^\mu_{~\nu}(k/M)\hat n^\nu(\tau)$ pointing in the direction of the particle spin where $\mathcal B^\mu_{~\nu}(k/M)$ is a boost from the particle rest frame to the computational frame, and $\hat n(\tau)$ is a unit 3-dimensional vector pointing in the direction of the particle spin in its rest frame with $\tau$ being the proper time.

\section{First order: the quantum correction}
In this section, we investigate the components of $G_{(1)}$ and neglect all the terms of the order $\hbar^2$. Eq.(\ref{eq:1scalar}-\ref{eq:2spin}) 
are reduced to
\begin{eqnarray}
\label{eq:1scalar1}
0 &=& D_\mu (V_{(0)}^\mu+\hbar V_{(1)}^\mu)- \partial_{X^\mu}\Phi\partial_{k_\mu}(S_{(0)}+\hbar S_{(1)})\nonumber\\
&& + \mathcal O(\hbar^2), \\
\label{eq:1pscalar1}
0 &=& 2\hbar A_{(1)}^\mu k_\mu + \mathcal O(\hbar^2), \\
\label{eq:1vector1}
0 &=& \hbar D_\mu S_{(0)} + 2\hbar k^\nu J^{(1)}_{\nu\mu} -\hbar \partial_{X^\nu}\Phi\partial_{k_\nu} V^{(0)}_\mu \nonumber\\ &&
+ \mathcal O(\hbar^2),\\
\label{eq:1avector1}
0 &=& \frac \hbar 2 \epsilon^{\mu\nu\rho\sigma} D_\mu J^{(0)}_{\rho\sigma} - 2\hbar k^\nu P_{(1)} +\hbar\partial_{X^\mu}\Phi\partial_{k_\mu} A_{(0)}^\nu \nonumber\\
&& + \mathcal O(\hbar^2), \\
\label{eq:1spin1}
0 &=& \frac \hbar 2 \epsilon^{\mu\nu\rho\sigma} D_\mu A^{(0)}_\nu - \hbar (k^\rho V^\sigma_{(1)} - k^\sigma V^\rho_{(1)})\nonumber\\
&&-\frac \hbar 2 \partial_{X^\mu}\Phi\partial_{k_\mu} J_{(0)}^{\rho\sigma} + \mathcal O(\hbar^2), \\
\label{eq:2scalar1}
0 &=& \hbar k_\mu V^\mu_{(1)} -\hbar MS_{(1)} + \mathcal O(\hbar^2), \\
\label{eq:2pscalar1}
0 &=& \hbar D_\mu A^\mu_{(0)} - 2\hbar MP_{(1)} + \mathcal O(\hbar^2),\\
\label{eq:2vector1}
0 &=& \hbar D_\mu J^{\nu\mu}_{(0)} + 2\hbar k^\nu S_{(1)} - 2\hbar MV^\nu_{(1)} + \mathcal O(\hbar^2),\\
\label{eq:2avector1}
0 &=& \hbar \epsilon_{\rho\sigma\nu\mu} J^{\rho\sigma}_{(1)}k^\nu + 2\hbar MA_\mu^{(1)}  + \mathcal O(\hbar^2), \\
\label{eq:2spin1}
0 &=& \hbar D_\mu V_\nu^{(0)} - \hbar D_\nu V_\mu^{(0)} +2 \hbar \epsilon_{\mu\nu\rho\sigma}k^\rho A^\sigma_{(1)}\nonumber\\
&&-2\hbar MJ_{\mu\nu}^{(1)}+ \mathcal O(\hbar^2).
\end{eqnarray}
Among the above, Eq. (\ref{eq:1pscalar1}) is redundant with Eq.(\ref{eq:2avector1}), Eq.(\ref{eq:1avector1}) and Eq.(\ref{eq:2pscalar1}) describe the spin precession of the quark, and Eq.(\ref{eq:1scalar1}), Eq.(\ref{eq:2scalar1}) and Eq.(\ref{eq:2vector1}) give the quantum correction to the classical transport equation.
\subsection{spin precession}
Substituting $A_{(0)}$ and $J_{(0)}$ with the right hand sides of Eq.(\ref{eq:A0}) and Eq.(\ref{eq:J0}), and using $\epsilon_{\mu\nu\alpha\beta}\epsilon^{\mu\nu\rho\sigma} = - 2(\delta^\rho_\alpha \delta^\sigma_\beta - \delta^\rho_\beta\delta^\sigma_\alpha)$, we obtain from Eq.(\ref{eq:1avector1}) and Eq.(\ref{eq:2pscalar1}) that
\begin{eqnarray}
     0 &=& D_\mu(k^\nu N^\mu - k^\mu N^\nu)-k^\nu D_\mu N^\mu + \frac{k^\nu}{M} N^\mu \partial_{X^\mu}\Phi \nonumber\\
     && + M \partial_{X^\mu}\Phi \partial_{k_\mu} N^\nu + \mathcal O(\hbar),
\end{eqnarray}
where $N^\mu \doteq n^\mu(X,k)\sum_s s f^{(0)}_s(X,k)\delta(k^2-M^2)$. We further simplify the above equation, using Eq.(\ref{eq:Dk}), as
\begin{equation}
\label{eq:spinflow}
    (k^\mu D_\mu - M \partial_{X^\mu}\Phi\partial_{k_\mu})N^\nu = (\frac{k^\nu}{M}\partial_{X_\mu} \Phi - F^{\nu\mu})N_\mu + \mathcal O(\hbar).
\end{equation}
The derivatives on $f^{(0)}_s\delta(k^2-M^2)$ in Eq.(\ref{eq:spinflow}) vanish due to Eq.(\ref{eq:ClassTransport}) if the semi-classical transport equations are assumed to be symmetric about $s$. We therefore obtain the equation describing the motion of $n$, i.e.,
\begin{equation}
    \label{eq:precession}
    (k^\mu D_\mu - M \partial_{X^\mu}\Phi\partial_{k_\mu})n^\nu = (\frac{k^\nu}{M}\partial_{X_\mu} \Phi - F^{\nu\mu})n_\mu + \mathcal O(\hbar).
\end{equation}
To have a better understanding of Eq.(\ref{eq:precession}), let us look at the spin precession of a slowly moving quark. For a slowly moving quark, the boosting from its rest frame can be written as $\mathcal B \approx I - \zeta \hat{\mathbf k} \cdot \mathbf K$ where $\zeta \doteq \frac 1 2 \ln \frac{k^0 + |\mathbf k|}{k^0 - |\mathbf k|}$ is the rapidity and $(K^i)^\mu_{~\nu} \doteq \delta^{i\mu}\delta^0_\nu + \delta^{0\mu}\delta^i_\nu$ are the boost generators, and its derivative at a small $|\mathbf k|$ limit is $\lim_{|\mathbf k| \to 0} d\mathcal B = -\frac 1 M d\mathbf k \cdot \mathbf K$. In such a simple case, we can write down the equations describing the motion of $\hat n$, i.e.,
\begin{eqnarray}
\dot{\hat n}^0 &=& 0,\\
M \dot{\hat n}^i &=& F_j^{~i}\hat n ^j.
\end{eqnarray}
As expected, the time component of $\hat n$ should always be zero, and interestingly, the spin precession in the particle rest frame is dominated by the vector force while the scalar force plays no role.
\subsection{quantum correction to the transport equation}
Let us decompose $V_{(1)}^\mu$ into the components parallel and perpendicular to $k$, i.e.,
\begin{equation}
\label{eq:V1}
    V^\mu_{(1)} = k^\mu \mathcal F_{(1)}(X,k) + H^\mu (X,k)
\end{equation}
 where $k\cdot H = 0$. We learn from Eq.(\ref{eq:2scalar1}) that
\begin{equation}
\label{eq:S1}
  S_{(1)} = k^2 \mathcal F_{(1)}/M + \mathcal O(\hbar).   
\end{equation}
Substituting $V_{(1)}$ and $S_{(1)}$ in Eq.(\ref{eq:2vector1}) with the right hand sides of Eq.(\ref{eq:V1}) and Eq.(\ref{eq:S1}), we obtain
\begin{equation}
\label{eq:J0F1H}
    M D_\nu J^{\mu\nu}_{(0)} + 2 k^\mu (k^2 - M^2) \mathcal F_{(1)} - 2 M^2H^\mu + \mathcal O(\hbar) = 0.
\end{equation}
We first solve $\mathcal F_{(1)}$ by multiplying $k_\mu$ on both the sides of Eq.(\ref{eq:J0F1H}), and the solution is
\begin{equation}
\label{eq:F1a}
    \mathcal F_{(1)} = -\frac{M k_\mu D_\nu J^{\mu\nu}_{(0)}}{2 k^2 (k^2 - M^2)} + \mathcal O(\hbar).
\end{equation}
We obtain from Eq.(\ref{eq:J0}) that
\begin{eqnarray}
\label{eq:DJ0}
D_\nu J^{\mu\nu}_{(0)} &=& -\sum_{s=\pm} s\epsilon^{\mu\nu\rho\sigma}D_\nu [k_\rho n_\sigma f_s^{(0)}\delta(k^2-M^2)] \nonumber\\
&=& -2\widetilde{F}^{\mu\sigma}N_\sigma 
-\epsilon^{\mu\nu\rho\sigma}k_\rho D_\nu N_\sigma, 
\end{eqnarray}
where the second term vanishes after being multiplied by $k_\mu$ and $\widetilde{F}^{\mu\nu} \doteq \frac 1 2 \epsilon^{\mu\nu\rho\sigma} F_{\rho\sigma}$.
Therefore,
\begin{equation}
\label{eq:F1b}
    \mathcal F_{(1)} = \sum_{s=\pm} s\frac{M}{k^2}\widetilde{F}^{\mu\nu}k_\mu n_\nu f_s^{(0)} \frac{\delta(k^2-M^2)}{(k^2 - M^2)} + \mathcal O(\hbar).
\end{equation}
Using
\begin{eqnarray}
\label{eq:Ddelta}
x\delta^\prime(x) &=& -\delta(x), \nonumber\\
x\delta^{\prime\prime}(x) &=& -2 \delta^\prime(x),
\end{eqnarray}
we further simplify Eq.(\ref{eq:F1b}) as
\begin{equation}
\label{eq:F1c}
    \mathcal F_{(1)} = -\sum_{s=\pm} s\frac{M}{k^2}\widetilde{F}^{\mu\nu}k_\mu n_\nu f_s^{(0)} \delta^\prime(k^2-M^2) + \mathcal O(\hbar),
\end{equation}
which is an off-shell correction to $\mathcal F_{(0)}$. So
\begin{equation}
\label{eq:S1a}
    \mathcal S_{(1)} = -\sum_{s=\pm} s\widetilde{F}^{\rho\sigma}k_\rho n_\sigma f_s^{(0)} \delta^\prime(k^2-M^2) + \mathcal O(\hbar).
\end{equation}
We then substitute $D_\nu J^{\mu\nu}$ and $S_{(1)}$ in Eq.(\ref{eq:2vector1}) with the right hand sides of Eq.(\ref{eq:DJ0}) and Eq.(\ref{eq:S1a}) and obtain that
\begin{widetext}
\begin{eqnarray}
\label{eq:V1a}
V^\mu_{(1)} &=& \frac 1 {2M} \left[-2\widetilde{F}^{\mu\sigma}N_\sigma-\epsilon^{\mu\nu\rho\sigma}k_\rho D_\nu N_\sigma - 2 \sum_{s=\pm} s k^\mu \widetilde{F}^{\rho\sigma}k_\rho n_\sigma f_s^{(0)} \delta^\prime(k^2-M^2) \right]+ \mathcal O(\hbar) \nonumber\\
&=& \sum_{s=\pm} \frac s {2M} \left[\left(-2\widetilde{F}^{\mu\sigma}n_\sigma f_s^{(0)}-\epsilon^{\mu\nu\rho\sigma}k_\rho D_\nu (n_\sigma f_s^{(0)})\right)\delta(k^2-M^2) \right. \nonumber\\
&& \left. + 2\left(- k^\mu \widetilde{F}^{\rho\sigma}k_\rho n_\sigma-\epsilon^{\mu\nu\rho\sigma}k_\rho n_\sigma(F_{\nu\tau}k^\tau+M\partial_{X^\nu}\Phi)\right) f_s^{(0)} \delta^\prime(k^2-M^2) \right] + \mathcal O(\hbar),
\end{eqnarray}
where $-k^\tau\epsilon^{\mu\nu\rho\sigma}k_\rho n_\sigma F_{\nu\tau}$ can be written, using the Schouten identity
\begin{equation}
\label{eq:Schouten}
    k^\tau\epsilon^{\mu\nu\rho\sigma} + k^\mu\epsilon^{\nu\rho\sigma\tau} + k^\nu\epsilon^{\rho\sigma\tau\mu} + k^\rho\epsilon^{\sigma\tau\mu\nu} + k^\sigma\epsilon^{\tau\mu\nu\rho} = 0,
\end{equation}
as
\begin{eqnarray}
\label{eq:ekknF}
-k^\tau\epsilon^{\mu\nu\rho\sigma}k_\rho n_\sigma F_{\nu\tau} &=& 2k^\mu\widetilde F^{\rho\sigma}k_\rho n_\sigma + k^\tau\epsilon^{\rho\sigma\mu\nu}k_\rho n_\sigma F_{\nu\tau} + 2k^2 \widetilde F ^{\sigma\mu} n_\sigma + 2(k\cdot n) \widetilde F^{\mu\rho}k_\rho \nonumber\\
&=& k^\mu\widetilde F^{\rho\sigma}k_\rho n_\sigma + k^2 \widetilde F ^{\sigma\mu} n_\sigma.
\end{eqnarray}
Eq.(\ref{eq:V1a}) can therefore be simplified as
\begin{eqnarray}
\label{eq:V1b}
V^\mu_{(1)} 
&=& \sum_{s=\pm} \frac s {2M} \left[\left(-2\widetilde{F}^{\mu\sigma}n_\sigma f_s^{(0)}-\epsilon^{\mu\nu\rho\sigma}k_\rho D_\nu (n_\sigma f_s^{(0)})\right)\delta(k^2-M^2) \right. \nonumber\\
&& \left. + 2\left( k^2 \widetilde F ^{\sigma\mu} n_\sigma-\epsilon^{\mu\nu\rho\sigma}k_\rho n_\sigma M\partial_{X^\nu}\Phi\right) f_s^{(0)} \delta^\prime(k^2-M^2) \right] + \mathcal O(\hbar).
\end{eqnarray}
After adding all the ingredients, i.e., $S_{0}$, $V_{(0)}$, $S_{(1)}$ and $V_{(1)}$ in Eq.(\ref{eq:1scalar1}), we must have a transport equation written as the superposition of the terms proportional to $\delta (k^2-M^2)$, $\delta^\prime (k^2-M^2)$ and $\delta^{\prime\prime} (k^2-M^2)$. The term proportional to $\delta^{\prime\prime}(k^2-M^2)$ must come from the derivatives on $\delta^\prime(k^2-M^2)$, i.e.,
\begin{eqnarray}
\label{eq:delta''a}
&&\sum_{s=\pm} s\hbar\left( \frac{k^2}{M} \widetilde F ^{\sigma\mu} n_\sigma-\epsilon^{\mu\nu\rho\sigma}k_\rho n_\sigma \partial_{X^\nu}\Phi\right) f_s^{(0)} D_\mu\delta^\prime(k^2-M^2)+\sum_{s=\pm} s\hbar\widetilde{F}^{\rho\sigma}k_\rho n_\sigma f_s^{(0)}\partial_{X^\mu}\Phi\partial_{k_\mu}\delta^\prime(k^2-M^2) + \mathcal O(\hbar^2)\nonumber\\
&=& 2\sum_{s=\pm} s\hbar\left( \frac{k^2}{M} \widetilde F ^{\sigma\mu} n_\sigma-\epsilon^{\mu\nu\rho\sigma}k_\rho n_\sigma \partial_{X^\nu}\Phi\right) (F_{\mu\tau}k^\tau+M\partial_{X^\mu}\Phi)f_s^{(0)}\delta^{\prime\prime}(k^2-M^2)\nonumber\\
&& +2\sum_{s=\pm} s\hbar\widetilde{F}^{\rho\sigma}k_\rho n_\sigma \partial_{X^\mu}\Phi k^\mu  f_s^{(0)}\delta^{\prime\prime}(k^2-M^2) + \mathcal O (\hbar^2)
\end{eqnarray}
It is apparent that the term proportional to $\epsilon^{\mu\nu\rho\sigma}\partial_{X^\nu}\Phi\partial_{X^\mu}\Phi$ in Eq.(\ref{eq:delta''a}) vanishes. The term proportional to $\widetilde F ^{\sigma\mu} n_\sigma F_{\mu\tau}k^\tau$ vanishes as well since, using Eq.(\ref{eq:Schouten}) again, one can show that 
\begin{equation}
\label{eq:FFnk}
    \widetilde F ^{\sigma\mu} n_\sigma F_{\mu\tau}k^\tau = -\frac 1 4 (n\cdot k) \widetilde F^{\mu\nu} F_{\mu\nu} = 0.
\end{equation} The remaining terms in Eq.(\ref{eq:delta''a}) are
\begin{eqnarray}
\label{eq:delta''b}
2\sum_{s=\pm} s\hbar\left( k^2 \widetilde F ^{\sigma\nu} n_\sigma + \widetilde{F}^{\rho\sigma}k_\rho n_\sigma k^\nu-\epsilon^{\mu\nu\rho\sigma}k_\rho n_\sigma F_{\mu\tau}k^\tau\right) \partial_{X^\nu}\Phi f_s^{(0)}\delta^{\prime\prime}(k^2-M^2) + \mathcal O(\hbar^2),
\end{eqnarray}
which, according to Eq.(\ref{eq:ekknF}), is equal to $\mathcal O(\hbar^2)$. So there is no term, up to the order of $\hbar$, proportional to $\delta^{\prime\prime}(k^2 - M^2)$ in the transport equation.

The term proportional to $\delta^\prime(k^2-M^2)$ is
\begin{eqnarray}
\label{eq:delta'a}
&& \sum_{s=\pm} s\hbar D_\mu\left( \frac{k^2}{M} \widetilde F ^{\sigma\mu} n_\sigma f_s^{(0)}-\epsilon^{\mu\nu\rho\sigma}k_\rho n_\sigma f_s^{(0)} \partial_{X^\nu}\Phi\right) \delta^\prime(k^2-M^2)+\sum_{s=\pm} s\hbar\partial_{X^\mu}\Phi\partial_{k_\mu}\left(\widetilde{F}^{\rho\sigma}k_\rho n_\sigma f_s^{(0)}\right)\delta^\prime(k^2-M^2)\nonumber\\
&& + \sum_{s=\pm} \frac {s\hbar} {2M} \left(-2\widetilde{F}^{\mu\sigma}n_\sigma f_s^{(0)}-\epsilon^{\mu\nu\rho\sigma}k_\rho D_\nu (n_\sigma f_s^{(0)})\right)D_\mu \delta(k^2-M^2) + \mathcal O(\hbar^2) \nonumber\\
&=& \sum_{s=\pm} s\hbar \left[\frac{k^2-M^2}{M^2} \partial_{X^\mu}\Phi\widetilde F ^{\sigma\mu} n_\sigma f_s^{(0)} + \widetilde{F}^{\rho\sigma}k_\rho \partial_{X^\nu}\Phi\partial_{k_\nu}(n_\sigma f_s^{(0)})\right] \delta^\prime(k^2-M^2)\nonumber\\
&&  + \sum_{s=\pm} \frac {s\hbar} {M} \left[k^2 \widetilde F ^{\sigma\mu} D_\mu (n_\sigma f_s^{(0)}) - k^\tau\epsilon^{\mu\nu\rho\sigma}k_\rho F_{\mu\tau} D_\nu (n_\sigma f_s^{(0)}) \right]\delta^\prime(k^2-M^2) + \mathcal O(\hbar^2)
\end{eqnarray}
The above equality is obtained using Eq.(\ref{eq:FFnk}) and the Bianchi identity $\partial_\mu \widetilde F^{\mu\nu} = 0$. And using Eq.(\ref{eq:Schouten}) agian, we further simplify Eq.(\ref{eq:delta'a}) as
\begin{eqnarray}
\label{eq:delta'b}
&& \sum_{s=\pm} s\hbar \left[\frac{k^2-M^2}{M^2} \partial_{X^\mu}\Phi\widetilde F ^{\sigma\mu} n_\sigma f_s^{(0)} + \widetilde{F}^{\rho\sigma}k_\rho \partial_{X^\nu}\Phi\partial_{k_\nu}(n_\sigma f_s^{(0)})\right] \delta^\prime(k^2-M^2)\nonumber\\
&&  - \sum_{s=\pm} \frac {s\hbar} {M} \left[\widetilde F^{\rho\sigma} k_\rho k^\mu D_\mu (n_\sigma f_s^{(0)}) +  \widetilde F^{\mu\rho} k_\rho k^\sigma D_\mu (n_\sigma f_s^{(0)}) \right]\delta^\prime(k^2-M^2) + \mathcal O(\hbar^2)
\end{eqnarray}
The last term of Eq.(\ref{eq:delta'b}) vanishes since 
\begin{eqnarray}
&& - \widetilde F^{\mu\rho} k_\rho k^\sigma D_\mu (n_\sigma f_s^{(0)}) =  D_\mu(\widetilde F^{\mu\rho} k_\rho k^\sigma) n_\sigma f_s^{(0)}\nonumber\\
&=& \partial_\mu \widetilde F^{\mu\rho} k_\rho k^\sigma n_\sigma + (k \cdot n)\widetilde F^{\mu\rho} F_{\mu\rho} f_s^{(0)} + \widetilde F^{\mu\rho} k_\rho F_{\mu\sigma} n^\sigma f_s^{(0)},
\end{eqnarray}
where the first term vanishes due to the Bianchi identity, the second terms vanishes due to $n\cdot k = 0$ and the third term vanishes due to Eq.(\ref{eq:FFnk}). The remaining terms of Eq.(\ref{eq:delta'b}) can be re-organized as
\begin{eqnarray}
\label{eq:delta'c}
 \sum_{s=\pm} s\hbar \left[\frac{k^2-M^2}{M^2} \partial_{X^\mu}\Phi\widetilde F ^{\sigma\mu} n_\sigma f_s^{(0)} -\frac 1 M \widetilde{F}^{\rho\sigma}k_\rho \left(k^\mu D_\mu - M\partial_{X^\nu}\Phi\partial_{k_\nu}\right)(n_\sigma f_s^{(0)})\right] \delta^\prime(k^2-M^2) + \mathcal O(\hbar^2),
\end{eqnarray}
where the derivatives on $n_\sigma$ in the second term can be calculated in advance using Eq.(\ref{eq:precession}) and the terms including these derivatives, equal to $-\sum_s\frac s M \widetilde{F}^{\rho\sigma}k_\rho (\frac{k_\sigma}{M}\partial_{X^\mu} \Phi - F_{\sigma\mu})(n^\mu f_s^{(0)})$, vanish due to Eq.(\ref{eq:FFnk}). Eq.(\ref{eq:delta'c}) can finally be simplified as
\begin{eqnarray}
\label{eq:delta'd}
 && \sum_{s=\pm} s\hbar \left[\frac{k^2-M^2}{M^2} \partial_{X^\mu}\Phi\widetilde F ^{\sigma\mu} n_\sigma f_s^{(0)} -\frac 1 M \widetilde{F}^{\rho\sigma}k_\rho n_\sigma (k^\mu D_\mu - M\partial_{X^\nu}\Phi\partial_{k_\nu})f_s^{(0)}\right] \delta^\prime(k^2-M^2) + \mathcal O (\hbar^2),\nonumber\\
 &=& - \sum_{s=\pm} \left[\frac{s\hbar}{M^2} \partial_{X^\mu}\Phi\widetilde F ^{\sigma\mu} n_\sigma f_s^{(0)}\right] \delta(k^2-M^2) - \sum_{s=\pm} \left[\frac {s\hbar} M \widetilde{F}^{\rho\sigma}k_\rho n_\sigma (k^\mu D_\mu - M\partial_{X^\nu}\Phi\partial_{k_\nu})f_s^{(0)}\right] \delta^\prime(k^2-M^2)+ \mathcal O (\hbar^2),\nonumber\\
\end{eqnarray}
where the equality holds due to Eq.(\ref{eq:Ddelta}).

We finally write down the terms proportional to $\delta(k^2 - M^2)$, which include the contribution from Eq.(\ref{eq:delta'd}) as well, and they are
\begin{eqnarray}
\label{eq:onshell}
&&\sum_{s=\pm} \left\{
k^\mu D_\mu f^{(0)}_s - M\partial_{X^\mu}\Phi\partial_{k_\mu}f^{(0)}_s
- \frac{s\hbar}{M^2} \partial_{X^\mu}\Phi\widetilde F ^{\sigma\mu} n_\sigma f_s^{(0)} \right. \nonumber\\
&&\left. + D_\mu \left[\frac {s\hbar} {2M} \left(-2\widetilde{F}^{\mu\sigma}n_\sigma f_s^{(0)}-\epsilon^{\mu\nu\rho\sigma}k_\rho D_\nu (n_\sigma f_s^{(0)})\right)\right]
\right\} \delta(k^2-M^2) + \mathcal O(\hbar^2) \nonumber\\
&=&\sum_{s=\pm} \left\{
k^\mu D_\mu f^{(0)}_s - M\partial_{X^\mu}\Phi\partial_{k_\mu}f^{(0)}_s - \frac {s\hbar} {2M^2} \epsilon^{\mu\nu\rho\sigma}k_\rho \partial_{X^\mu}\Phi D_\nu (n_\sigma f_s^{(0)}) \right. \nonumber\\
&&\left. -  \frac {s\hbar} {2M}
\epsilon^{\mu\nu\rho\sigma}k_\rho D_\mu D_\nu (n_\sigma f_s^{(0)})
\right\} \delta(k^2-M^2)  + \mathcal O(\hbar^2) \nonumber\\
&=&\sum_{s=\pm} \left\{
k^\mu D_\mu f^{(0)}_s - M\partial_{X^\mu}\Phi\partial_{k_\mu}f^{(0)}_s - \frac {s\hbar} {2M^2} \epsilon^{\mu\nu\rho\sigma}k_\rho \partial_{X^\mu}\Phi D_\nu (n_\sigma f_s^{(0)}) \right. \nonumber\\
&&\left. +  \frac {s\hbar} {4M}
\epsilon^{\mu\nu\rho\sigma}k_\rho \partial_{X^\tau} F_{\mu\nu} \partial_{k_\tau} (n_\sigma f_s^{(0)})
\right\} \delta(k^2-M^2)  + \mathcal O(\hbar^2),
\end{eqnarray}
where the first equality is obtained using the Bianchi identity and the second one is obtained using Eq.(\ref{eq:DD}). Combining both Eq.(\ref{eq:delta'd}) and Eq.(\ref{eq:onshell}), and using the Tylor expansion, i.e., $\delta(x + \hbar y) = \delta(x) + \hbar y \delta^\prime(x) + \mathcal O(\hbar^2)$, we write down the transport equation as
\begin{eqnarray}
\label{eq:Qtransport}
0 &=& \sum_{s=\pm}\left[
k^\mu D_\mu f^{(0)}_s - M\partial_{X^\mu}\Phi\partial_{k_\mu}f^{(0)}_s - \frac {s\hbar} {2M^2} \epsilon^{\mu\nu\rho\sigma}k_\rho \partial_{X^\mu}\Phi D_\nu (n_\sigma f_s^{(0)})  +  \frac {s\hbar} {2M}
k_\rho \partial_{X^\tau} \widetilde F^{\rho\sigma} \partial_{k_\tau} (n_\sigma f_s^{(0)})
\right]\nonumber\\
&& \times \delta(k^2-M^2-\frac {s\hbar} M \widetilde{F}^{\rho\sigma}k_\rho n_\sigma) + \mathcal O(\hbar^2).
\end{eqnarray}

It seems that we have already obtained the Vlasov equation for the spinning quarks. But before drawing conclusions, let's take a second look at the physical meaning of $f^{(0)}_s$ by writing down the current density up to the order of $\hbar$, which is
\begin{eqnarray}
\label{eq:current}
j^\mu &=& \int d^4k (V^\mu_{(0)}+\hbar V^\mu_{(1)}) 
= \sum_{s=\pm} \int d^4k \left\{\left[k^\mu f_s^{(0)} +\frac {s\hbar} {2M}\left(-2\widetilde{F}^{\mu\sigma}n_\sigma f_s^{(0)}-\epsilon^{\mu\nu\rho\sigma}k_\rho D_\nu (n_\sigma f_s^{(0)})\right)\right]\delta(k^2-M^2) \right. \nonumber\\
&& \left. + \frac {s\hbar} {M}\left( k^2 \widetilde F ^{\sigma\mu} n_\sigma-\epsilon^{\mu\nu\rho\sigma}k_\rho n_\sigma M\partial_{X^\nu}\Phi\right) f_s^{(0)} \delta^\prime(k^2-M^2) \right\} + \mathcal O(\hbar^2).
\end{eqnarray}
A bizarre point is that even if $f_s^{(0)}$ is spatially uniform and isotropic in the momentum space, $\vec j$ is not vanishing, rather it is, if the off-shell contribution is neglected, equal to
\begin{equation}
    j^i = -\sum_{s=\pm} \int d^4k \frac {s\hbar} {M}\left(\widetilde{F}^{i\sigma}n_\sigma f_s^{(0)}\right)\delta(k^2-M^2)= - \frac \hbar {M^2} \widetilde{F}^{i\sigma}j_{5\sigma}.
\end{equation}
It means that a small portion of the quark motion is not included in $f^{(0)}_s$ directly, which might not be a defect, since one can always argue that although $f^{(0)}_s$ is no longer the precise distribution of the spinning quarks at the order of $\hbar$, it could be the distribution of some quasi-particles from which one can calculate all the macroscopic currents. But it is always better if the quark motions could be described by $f_s$ directly, and it can be achieved if we make such a substitution that $k^\mu \to \kappa^\mu +\frac{s\hbar}{M}\widetilde{F}^{\mu\sigma}n_\sigma$, $\partial_{k_\mu} \to \partial_{\kappa_\mu} - \frac{s\hbar}{M}\widetilde F_{\nu\sigma} \partial_{\kappa_\mu} n^\sigma \partial_{\kappa_\nu} + \mathcal O(\hbar^2)$, $\partial_{X^\mu} \to \partial_{X^\mu} - s\hbar \partial_{X^\mu}(\widetilde F_{\nu\sigma}\frac {n^\sigma} M)\partial_{\kappa_\nu}$, $D_\mu \to D_\mu - s\hbar D_\mu(\widetilde F_{\nu\sigma}\frac {n^\sigma} M)\partial_{\kappa_\nu}$ and $f^{(0)}_s \to f_s {\rm det}|\partial \kappa / \partial k| = f_s(1-\frac{s\hbar} M \widetilde F^{\mu\sigma} \partial_{\kappa^\mu} n^\sigma) + \mathcal O(\hbar^2)$ where $\kappa$ fulfills $\kappa\cdot n = 0$. Under such a substitution, the current density is expressed as
\begin{eqnarray}
\label{eq:current1}
j^\mu &=& \sum_{s=\pm} \int d^4\kappa \left\{\left[\kappa^\mu f_s +\frac {s\hbar} {2M}\left(-\epsilon^{\mu\nu\rho\sigma}\kappa_\rho D_\nu (n_\sigma f_s)\right)\right]\delta(\kappa^2-M^2 + 2\frac {s\hbar} M \widetilde{F}^{\rho\sigma}\kappa_\rho n_\sigma) \right. \nonumber\\
&& \left. + \frac {s\hbar} {M}\left( \kappa^2 \widetilde F ^{\sigma\mu} n_\sigma-\epsilon^{\mu\nu\rho\sigma}\kappa_\rho n_\sigma M\partial_{X^\nu}\Phi\right) f_s \delta^\prime(\kappa^2-M^2+ 2\frac {s\hbar} M \widetilde{F}^{\rho\sigma}\kappa_\rho n_\sigma) \right\} + \mathcal O(\hbar^2)
\end{eqnarray}
whose spatial component vanishes if $f_s$ is spatially uniform and isotropic in $\kappa$. The transport equation Eq.(\ref{eq:Qtransport}) is therefore written as
\begin{eqnarray}
\label{eq:Qtransport2}
0 &=& \sum_{s=\pm}\left\{
\left(\kappa^\mu + \frac{s\hbar}{M}\widetilde{F}^{\mu\sigma}n_\sigma\right)D_\mu f_s  - M\partial_{X^\mu}\Phi\partial_{\kappa_\mu}f_s   -\frac{s\hbar}{M} \left[\kappa^\mu n^\sigma \partial_{X^\mu}\widetilde F_{\nu\sigma}+\widetilde F_{\nu\sigma}\left(\frac{\partial_{X^\mu}\Phi}{M} (\kappa^\sigma n^\mu+\kappa^\mu n^\sigma)-F^{\sigma\mu}n_\mu\right)\right] \partial_{\kappa_\nu} f_s\right.\nonumber\\
&&\left.- \frac {s\hbar} {2M^2} \epsilon^{\mu\nu\rho\sigma}\kappa_\rho \partial_{X^\mu}\Phi D_\nu (n_\sigma f_s)  +  \frac {s\hbar} {2M}
\kappa_\rho \partial_{X^\tau} \widetilde F^{\rho\sigma} \partial_{\kappa_\tau} (n_\sigma f_s)
\right\}
\delta\left(\kappa^2-M^2+\frac {s\hbar} M \widetilde{F}^{\rho\sigma}\kappa_\rho n_\sigma\right) + \mathcal O(\hbar^2).
\end{eqnarray}

The equations of motion of the particle can be obtained by replacing the distribution function $f_s$ in Eq.(\ref{eq:Qtransport2}) with $\sum_a \delta(\mathbf x - \mathbf x_a(t)) \delta(\boldsymbol \kappa - \boldsymbol \kappa_a(t))\delta(n-n_a)\delta_{ss_a}$ where $\mathbf x_a$, $\boldsymbol \kappa_a$, $s_a$ and $n_a$ are the position, momentum, spin and spin direction of the particle respectively, and the equations are
\begin{eqnarray}
\label{eq:eomx}
 \dot{\mathbf x}_a &=& \frac{\boldsymbol \kappa_a + \frac{s_a\hbar}{M}\left[\mathbf Y_a+\frac 1 {2M} \left(\partial_t\Phi \mathbf n_a \times \boldsymbol \kappa + k^0_a \nabla \Phi \times \mathbf n_a + n_a^0 \boldsymbol \kappa \times \nabla \Phi\right)\right]}
 {\kappa^0_a - \frac{s\hbar}{M}\mathbf B \cdot \mathbf n_a + \frac{s_a\hbar}{2M^2} \nabla \Phi \cdot (\boldsymbol \kappa_a \times \mathbf n_a)}, \\
\label{eq:eomk}
 \dot{\boldsymbol \kappa}_a &=& \frac{M\nabla \Phi + \frac{s_a\hbar}{M}\left[(\mathbf E\cdot \mathbf B) \mathbf n_a + \frac{\partial_{X^\mu}\Phi}{M} \left(n^\mu_a ( \kappa^0_a \mathbf B- \boldsymbol \kappa_a \times \mathbf E) - \kappa^\mu_a \mathbf Y_a\right) + \kappa^\mu_a \partial_{X^\mu}\mathbf Y_a-\frac 1 2 \kappa_\mu n^a_\nu \nabla \widetilde F^{\mu\nu} \right]}
 {\kappa^0_a - \frac{s\hbar}{M}\mathbf B \cdot \mathbf n_a + \frac{s_a\hbar}{2M^2} \nabla \Phi \cdot (\boldsymbol \kappa_a \times \mathbf n_a)}\nonumber\\
 &&+ \mathbf E - \dot{\mathbf x}_a \times \mathbf B,
\end{eqnarray}
\end{widetext}
where $\mathbf E \doteq \partial_t \mathbf \Omega + \nabla \Omega^0$, $\mathbf B \doteq \nabla \times \mathbf \Omega$, $\mathbf Y_a \doteq \mathbf n_a \times \mathbf E -n^0_a\mathbf B$ and $\kappa^0_a$ is constrained by $\kappa_a^2 - M^2 + \frac{s_a\hbar}{M}\widetilde F_{\rho\sigma} \kappa_a^\rho n_a^\sigma = 0$. It turns out that both the scalar and vector force gives rise to an anomalous velocity, while the anomalous force vanishes if the vector force is turned off. The dispersion relation is modified by the vector force as well.

\section{Conclusion and outlook}
We derive the transport equation and the equations of motion for the spinning quarks, which are moving in the NJL-type mean field, by solving the Dyson equations,  up to the order of $\hbar$, under the mean field approximation, and find that the scalar and the vector mean field potentials have different impacts on the quantum correction to both the classical transport equation and the equations of motion. The scalar force gives rise to an anomalous velocity and contributes, with the vector force, to the anomalous force which vanishes if the vector force is turned off, while the vector force gives rise to both an anomalous velocity and an anomalous force, and modifies the particle dispersion relation as well. Besides, the spin precession in the particle rest frame is dominated by the vector force. Therefore, the motions of the spinning quarks, in a baryon rich rotating fireball, can be very different in the cases with and without the vector mean field potential, which might provide a new perspective on studying the existence of the vector interactions among the quarks. We will derive the full transport equation including the collision terms in the future, and a simulation, where the motions of the particles are governed by Eq.(\ref{eq:eomx}) and Eq.(\ref{eq:eomk}), is undergoing.

\bibliography{references}

\end{document}